# Unified Model of Ferroelectricity Induced by Spin Order

H. J. Xiang[1*], P. S. Wang[1], M.-H. Whangbo[2], and X. G. Gong[1]

[1] Key Laboratory of Computational Physical Sciences (Ministry of Education), State Key Laboratory of Surface Physics, and Department of Physics, Fudan University, Shanghai 200433, P. R. China

[2]Department of Chemistry, North Carolina State University, Raleigh, North Carolina 27695-8204, USA

## Abstract

The ferroelectricity of multiferroics induced by spin order is commonly explained by considering either purely electronic or ion-displacement contribution. However, there is no general model which includes both effects simultaneously. Here, we suggest a realistic model to describe the ion-displacement part of the ferroelectricity based on the spin-lattice coupling Hamiltonian. Combining this model with our previous pure electronic model for spin-order induced polarization, we propose a unified model that includes both effects. By applying the unified model to representative multiferroics where the electronic and ion-displacement contributions vary widely, we find that this model can not only reproduce the first-principles results, but also provide insight into the origin of ferroelectricity.

PACS numbers: **75.85.+t, 71.15.Rf, 71.20.-b, 77.80.-e**

Multiferroics [1-4] have attracted much attention largely because coupling between ferromagnetism and ferroelectricity might lead to additional novel effects. A fundamental issue concerning multiferroics is the origin of their ferroelectric (FE) polarization, for which two different mechanisms have been considered; one is the pure electronic mechanism in which a spin order induces charge redistribution responsible for the polarization [5-8], and the other is the pure nuclear mechanism in which the polarization arises from ion displacements [9-11]. The electronic mechanism can be divided into two groups depending on whether or not spin-orbit coupling (SOC) is involved; the exchange striction mechanism ($\vec{P}_{ij} \propto \vec{S}_i \cdot \vec{S}_j$) does not depend on SOC [7,12], while the spin-current-like or the single-site mechanism arises from SOC [5,6,8,13]. The spin-current mechanism proposed by Katsura *et al.* [5], referred to as the KNB model, predicts $\vec{P}_{ij} \propto \vec{e}_{ij} \times (\vec{S}_i \times \vec{S}_j)$ and has been widely used to explain the FE phenomena of various multiferroics such as $TbMnO_3$, $MnWO_4$ and $LiCuVO_4$, which occurs as a consequence of spiral spin order. The general spin-order-induced polarization model [8,12,13] includes the exchange striction term [12], the general spin-current term [$\vec{P}_{ij} \propto \mathbf{M}_{ij}(\vec{S}_i \times \vec{S}_j)$] [8], and the single-site term [13], and explains the polarization induced not only by cycloidal and proper-screw spin spiral orders but also by a collinear ferrimagnetic spin order. Similarly, the nuclear mechanism can also be grouped into two categories; Sergienko and Dagotto [9] explained the ferroelectricity of $TbMnO_3$ in terms of ion displacements induced by inverse Dzyaloshinskii-Moria (DM) interaction (hence by SOC). The importance of ion displacements in the polarization of $TbMnO_3$ was also pointed out in the first principles studies [10,11]. The polarizations of some multiferroics [7,14,15] are caused by their ion displacements that occur to lower the Heisenberg exchange interaction energy and hence do not involve SOC. So far, there has been no general theory that can quantitatively describe the ion displacements leading to FE polarization. In addition, there has been no model unifying the pure electronic and pure nuclear contributions to FE polarization.

In this Letter, we develop a general method for describing the ion-displacement contribution to FE polarization on the basis of the spin-lattice interaction Hamiltonian, which does not require geometry relaxation, and then present a unified model that combines the pure electronic and the ion-displacement contributions. Then we test the unified model by investigating representative multiferroics $LiCuVO_4$, $TbMnO_3$ and $HoMnO_3$, in which the electronic and ion-displacement contributions vary widely, to find their ferroelectricity quantitatively described by the model.

In general, the total electric polarization $\vec{P}_t = \vec{P}_t(\vec{S}_1, \vec{S}_2, \cdots, \vec{S}_m;\ \vec{u}_1, \vec{u}_2, \cdots, \vec{u}_n)$ of a magnetic system is a function of the spin direction $\vec{S}_i$ of the magnetic ions $i$ and the displacement $\vec{u}_k$ of the ions $k$ (including non-magnetic ions) in the magnetic unit cell. Here the ion displacement $\vec{u}_k$ is given with respect to a reference structure, usually a paraelectric centrosymmetric structure. For simplicity, we use the notation $\vec{U} = (\vec{u}_1, \vec{u}_2, \cdots, \vec{u}_n)$ and $\vec{U} = 0$ when all $\vec{u}_k = 0$. In magnetic multiferroics, the ion displacement is rather small (usually $|\vec{u}_k| \ll 0.01$ Å), so that the total FE polarization can be estimated accurately by $\vec{P}_t \approx \vec{P}_e(\vec{S}_1, \vec{S}_2, \cdots, \vec{S}_m;\ \vec{U} = 0) + \vec{P}_{ion}(\vec{U})$, where $\vec{P}_e$ is the electronic contribution induced by a spin order, while $\vec{P}_{ion}$ is the ion-displacement contribution when the spins are in a paramagnetic (PM) state. We assume that the electronic contribution remains almost unchanged after tiny ion displacements, which we validate by direct density functional theory (DFT) calculations (see [16]). $\vec{P}_e$ consists of the single-site term, a symmetric exchange striction term, and the general spin-current term

$$\vec{P}_e = \sum_{i,\alpha\beta} \vec{P}_{i,\alpha\beta} S_{i\alpha} S_{i\beta} + \sum_{<i,j>} \vec{P}_{es}^{ij} \vec{S}_i \cdot \vec{S}_j + \sum_{<i,j>} \mathbf{M}^{ij} (\vec{S}_i \times \vec{S}_j),$$

where the single-site term $\vec{P}_{i,\alpha\beta}$ ($\alpha$, $\beta$ = $x$, $y$ or $z$) is usually small and can be neglected. The polarization matrix $\mathbf{M}^{ij}$ describes the generalized spin-current contribution

$$\mathbf{M}^{ij} = \begin{bmatrix} (\mathbf{P}_{ij}^{yz})_x & (\mathbf{P}_{ij}^{zx})_x & (\mathbf{P}_{ij}^{xy})_x \\ (\mathbf{P}_{ij}^{yz})_y & (\mathbf{P}_{ij}^{zx})_y & (\mathbf{P}_{ij}^{xy})_y \\ (\mathbf{P}_{ij}^{yz})_z & (\mathbf{P}_{ij}^{zx})_z & (\mathbf{P}_{ij}^{xy})_z \end{bmatrix}$$

where the element $(\mathbf{P}_{ij}^{\alpha\beta})_\gamma$ means the polarization along the γ (= x, y or z) direction when the spins at $i$ and $j$ are along the $\alpha$ and $\beta$ directions, respectively, and the $(\vec{S}_i \times \vec{S}_j)$ term is treated as a column vector [8,12,13].

In terms of the Born effective charges $Z_{i\alpha}$ and the displacements $u_{i\alpha}$ ( $\alpha$ = $x$, $y$ or $z$) of the ions $i$, the lattice contribution to the FE polarization is written as [17]

$$P_{ion,\alpha}(\vec{u}_1, \vec{u}_2, \cdots, \vec{u}_n) = \sum_i u_{i\alpha} Z_{i\alpha}.$$

We evaluate the ion displacements, caused by the forces associated with a spin order, by minimizing the spin-lattice interaction energy

$$E(\vec{S}_1, \vec{S}_2, \cdots, \vec{S}_m;\ \vec{U}) = E_{PM} + E_{ph} + E_{spin},$$

where $E_{PM}$ is the energy of the reference paraelectric PM state, and the elastic energy is given by $E_{ph} = \frac{1}{2}\sum_{ij\alpha\beta} C_{ij}^{\alpha\beta} u_{i\alpha} u_{j\beta}$ with $C_{kj}^{\alpha\beta}$ as the force constant. The spin interaction energy $E_{spin}$ can be written as

$$E_{spin} = \sum_{<i,j>} J_{ij}(\vec{U})\vec{S}_i \cdot \vec{S}_j + \vec{D}_{ij}(\vec{U}) \cdot (\vec{S}_i \times \vec{S}_j),$$

where $J_{ij}(\vec{U}) = J_{ij}^0 + \sum_{k\alpha} \frac{\partial J_{ij}}{\partial u_{k\alpha}} u_{k\alpha}$ is the symmetric exchange and $\vec{D}_{ij}(\vec{U})$ is the DM exchange [18,19] $\vec{D}_{ij}(U) = \vec{D}_{ij}^0 + \sum_{k\alpha} \frac{\partial \vec{D}_{ij}}{\partial u_{k\alpha}} u_{k\alpha}$. Both the symmetric exchange and the DM exchange depend on the ion displacements. The first order derivatives of these parameters can be efficiently evaluated by using the four-state mapping method [20,21]. Then, for any given spin order, the ion displacement can be obtained by solving the equation

$$\frac{\partial E}{\partial u_{k\alpha}} = \sum_{j\beta} C_{kj}^{\alpha\beta} u_{j\beta} + \frac{\partial E_{spin}}{\partial u_{k\alpha}} = 0$$

$$\text{with } \frac{\partial E_{spin}}{\partial u_{k\alpha}} = \sum_{<i,j>} \frac{\partial J_{ij}}{\partial u_{k\alpha}} \vec{S}_i \cdot \vec{S}_j + \frac{\partial \vec{D}_{ij}}{\partial u_{k\alpha}} \cdot (\vec{S}_i \times \vec{S}_j).$$

If one needs all the ion displacements, we need to solve these equations within a magnetic unit cell. Fortunately, for the purpose of calculating the polarization, the “average” displacements (i.e., $\Gamma$ phonon modes) within the chemical unit cell are sufficient. Thus, we first compute the “average” force for each ion of the chemical unit cell and then use the $\Gamma$ force constants to obtain the average ion displacements. This procedure is advantageous because it involves only linear equations with much smaller dimensions ($3N_{uc}$, where $N_{uc}$ is the number of ions in the chemical unit cell) and because we need only the force constants of a unit cell that can be calculated very easily. This approach, being linear-scaled, can deal with any magnetic state with very large magnetic unit cell. In principle, one prefers to use the $\Gamma$ force constant matrix and the Born effective charges corresponding to the PM state, which are difficult to obtain. To a first approximation, however, those of the ferromagnetic state are sufficient, which can be readily determined using the density functional perturbation theory [22,23] or finite difference method. In what follows, we test the unified model for three representative mutiferroics $LiCuVO_4$, $TbMnO_3$ and $HoMnO_3$, in which the electronic and ion-displacement contributions are widely different.

$LiCuVO_4$ [24] contains spin-frustrated $CuO_2$ ribbon chains [see Fig. 1(a)] made up of edge-sharing $CuO_4$ squares with spin-half magnetic ions $Cu^{2+}$. Because of the competition between nearest-neighbor (NN) ferromagnetic and next-nearest-neighbor antiferromagnetic interactions in each $CuO_2$ chain, a cycloidal spin spiral state sets in along the chain direction (i.e., the b direction) at low temperature [25]. We first consider the electronic contribution to the FE polarization. The NN Cu dimer with neighboring ligands has $C_{2v}$ (approximately local $D_{2h}$) point group symmetry. The symmetric exchange striction term is nonzero since there is no spatial inversion center between two NN Cu ions, but the net contribution of the symmetric exchange striction is zero in a spiral state. Thus the polarization induced by the spin spiral arises solely from the general spin-current term. With the coordination system defined in Fig. 2, where the y axis is along the Cu1 – Cu2 direction, the polarization matrix **M** (in units of $10^{-5}$ eÅ) obtained from the four-state mapping analysis [8] is given by

$$\mathbf{M}=\begin{bmatrix} 0 & 0 & -50.5 \\ 0 & 0 & 0 \\ -8.5 & 0 & 0 \end{bmatrix}.$$

The form of **M** is consistent with the local symmetry ($C_{2v}$) of the Cu dimer (see [16]). $M_{13} \equiv (\mathbf{P}^{xy})_x$ = -50.5 is much greater than $M_{31} \equiv (\mathbf{P}^{yz})_z$ = -8.5 in magnitude. (Note that the KNB model predicts $M_{13}$ = -$M_{31}$). The polarization is much greater for the *ab*-plane ($\alpha\beta$ = xy) than the *bc*-plane ($\alpha\beta$ = yz) spin spiral because the hole wave function (commonly referred to as $d_{x2-y2}$ but $d_{xy}$ in the local coordinate chosen) is contained in the *ab*-plane. From the general spin-current model, the electronic part of the polarization is calculated to be (-112.4, 0, 0) and (0, 0, 18.9) μC/m$^2$ for the *ab*- and *bc*-plane spin spiral, respectively, in good agreement with the corresponding results [(-103.5, 0, 0) and (0, 0, 15.7) μC/m$^2$] from direct DFT calculations [26].

To calculate the ion-displacement contribution to the polarization in $LiCuVO_4$, we extract the first order derivatives of the DM interaction between two NN Cu ions using the four-state mapping approach [20]. Our results show that the DM interaction for the experimental structure is almost zero, but its first-order derivative is not, as shown in Fig. 2; $\frac{\partial \vec{D}_{12}}{\partial u_{k\alpha}}$ is large for the two Cu ions and two bridging O ions with a non-negligible contribution from the neighboring V ions. These derivatives indicate that the cations ($Cu^{2+}$ and $V^{5+}$) and anions ($O^{2-}$) would have opposite displacements along z leading to nonzero $D_{12}^x$ [Fig. 2(a)], and similarly polar displacements along x will lead to nonzero $D_{12}^z$ [Fig. 2(b)]. From the polar ion-displacements obtained by using $\frac{\partial \vec{D}_{ij}}{\partial u_{k\alpha}}$, we find that the ion-displacement contribution to the polarization is (-366.9, 0, 0) μC/m$^2$ for the *ab*-plane spin spiral and (0, 0, 187.4) μC/m$^2$ for the *bc*-plane spin spiral. This contribution is larger for the *ab*-plane spin spiral because $|\frac{\partial D_{12}^{z}}{\partial u_{k\alpha}}|$ is larger than $|\frac{\partial D_{12}^{x}}{\partial u_{k\alpha}}|$. Note that the ion-displacement contribution to the polarization is in the same direction as the electronic contribution, in agreement with the results from the direct structural optimization [26]. The total polarization is also in good agreement with the direct DFT calculations [26], reflecting the accuracy of our unified model.

To gain insight into the nature of the ion-displacement contribution to the FE polarization in $LiCuVO_4$, we examine how the DM interaction depends on the ion displacements by considering $\frac{\partial D_{12}^{z}}{\partial u_{k\alpha}}$ as an example. As can be seen from Fig. 2(b), $D_{12}^{z}$ becomes large when Cu moves along the x direction and/or O move along the –x direction. We analyze this DFT result in terms of tight-binding (TB) calculations by considering the SOC/hopping interactions between two NN $Cu^{2+}$ ions. For simplicity, we consider the hole state $d_{X2-Y2}$ and the filled $d_{XY}$ state for each $Cu^{2+}$ site, as shown in Fig. 3, where the local coordinate system XYZ is adopted instead of the xyz coordinate system in Fig. 2. Our TB model employs a $Cu_2O_2$ cluster made up of two Cu ions and two O ions. If this cluster has $D_{2h}$ symmetry, no three-step hopping process is possible, thus there is no DM interaction [Fig. 3(a)]. If the two O ions move in the same direction perpendicular to the Cu-Cu direction (here, only the relative displacements between the $O^{2-}$ and $Cu^{2+}$ ions matter), the coupling between the neighboring $d_{X2-Y2}$ orbitals and that between the $d_{X2-Y2}$ orbital of one Cu ion and the $d_{XY}$ orbital of the neighboring Cu ion become non-zero. As a result, the DM interaction between Cu1 and Cu2 becomes nonzero [Figs. 3(b)]. The DM interaction parameter can be evaluated by using the perturbation theory (see [16]).

$TbMnO_3$, a classical spin-spiral multiferroic [27], has a distorted $GdFeO_3$-type orthorhombic perovskite structure with space group Pbnm [28]. It exhibits orbital ordering in the *ab*-plane [see Fig. 1(b)] where the long axial Mn-O bonds of one $MnO_6$ octahedron are connected by corner-sharing to the short Mn-O bonds of its adjacent $MnO_6$ octahedra. The spin order of $TbMnO_3$ exhibits a *bc*-plane spin spiral in the absence of magnetic field, and an *ab*-plane spin spiral in the presence of an in-plane magnetic field. There is no spatial inversion center between two NN in-plane Mn ions, but the total contribution from the symmetric exchange striction is zero in the perfect spiral state by symmetry. Using the method described for $LiCuVO_4$, we get the polarization matrix **M** (in units of $10^{-4}$ eÅ)

$$\mathbf{M} = \begin{bmatrix} 1.4 & -0.2 & -0.3 \\ 12.2 & -1.2 & -9.1 \\ -1.4 & 0.7 & -1.5 \end{bmatrix},$$

with a local coordinate system in which the y axis is along the Mn – Mn direction and the z axis is along *c*. The second-row elements of **M**, i.e., the y-components of the general polarization matrix elements **P**, are large so that the polarization for the Mn – Mn pair is almost along the Mn – Mn direction. This is due to the orbital ordering in the *ab*-plane of $TbMnO_3$, which leads to a strong interaction between the occupied $e_g$ level of one $Mn^{3+}$ ion and the unoccupied $e_g$ level of the neighboring $Mn^{3+}$ ion. As in the case of $LiCuVO_4$, the $M_{13}$ and $M_{31}$ elements are quite different, so that the KNB model is not supported as in the case of $LiCuVO_4$. In contrast, the general spin current model reproduces the DFT results very well, as can be seen from Table I. Note that the electronic contribution is large in the *ab*-plane spiral case, but very small for the *bc*-plane spiral. The direction of the polarization for the Mn – Mn pair for the *ab*-plane spiral case is almost perpendicular to the *c* axis, and has nonzero components along the *a*-direction.

These results are in good agreement with the previous DFT results [10,11]. This is because the symmetry analysis shows that the directions of the polarizations for the *ab*- and *bc*-plane spin spiral cases are along *a* and *c* axes, respectively. For the *ab*-plane spin spiral, the ion-displacement contribution is opposite in sign to the electronic contribution but is greater in magnitude. For the *bc*-plane spin spiral, the ion-displacement contribution is the same in sign as the electronic contribution but is much greater in magnitude. These findings are in complete accord with the direct DFT results (Table I) [10,11]. The small difference in magnitude between the DFT results and those from the unified model should be due to the fact only the NN Mn-Mn pair is taken into account in our current model analysis. The ion displacements play an important role in the total polarization, as pointed out in previous theoretical calculations and verified by subsequent experiments [29].

We now turn to the multiferroic $HoMnO_3$ for which the ferroelectricity is due to symmetric exchange striction. This multiferroic displays a collinear E-type AFM (E-AFM) spin order [see Fig. 1(b)] that breaks the spatial inversion symmetry. Due to the collinear spin arrangement, the generalized spin-current contribution is zero for $HoMnO_3$. Picozzi *et al.* showed that the electronic and the ion-displacement contributions are comparable in magnitude in $HoMnO_3$ [7]. By computing the coefficients $\vec{P}_{es}$ for the in-plane NN Mn – Mn exchange path, we obtain the electronic contribution (3.47 $\mu C/cm^2$), which is close to the direct DFT result (3.19 $\mu C/cm^2$). To obtain the ion-displacement contribution, we calculate the first derivative of the symmetry spin interaction for the in-plane NN Mn – Mn exchange paths. By solving the spin-lattice interaction Hamiltonian for the E-AFM order, the lattice contribution is found to be 3.28 $\mu C/cm^2$. The total polarization from our model is then 6.75 $\mu C/cm^2$, which nicely confirms the direct DFT result (6.14 $\mu C/cm^2$) [7]. As shown previously [7], the E-AFM spin order results in displacements of the $Mn^{3+}$ and $O^{2-}$ ions such that the Mn-O-Mn bond angle becomes smaller (larger) for the antiferromagnetic (ferromagnetic) Mn – Mn exchange path. This can be understood by considering the derivatives of the NN symmetric spin exchanges [16]: When the Mn-O-Mn angle becomes larger (closer to 180°), the spin exchange becomes more ferromagnetic because of the enhanced coupling between the occupied $d_{z2}$-like orbital of a Mn ion and the unoccupied $d_{x2-y2}$-like orbital of its neighboring Mn ion.

In summary, we developed an efficient method to describe the ion-displacement contribution to the spin-order-induced FE polarization. This allowed us to propose a model that unifies the electronic and ion-displacement contributions to the FE polarization for any spin order without performing additional DFT calculations. For three representative multiferroics in which the electronic and ion-displacement contributions are widely different, our unified model closely reproduces the results obtained directly from density functional calculations. It is expected to be applicable for a variety of multiferroics. Our model is not only useful for revealing the origin of ferroelectricity in multiferroics, but also useful for model studies on the magnetoelectric coupling.

Work at Fudan was partially supported by NSFC, the Special Funds for Major State Basic Research, FANEDD, Research Program of Shanghai municipality and MOE.

e-mail: hxiang@fudan.edu.cn

[1] S.-W. Cheong and M. Mostovoy, Nature Mater. **6**, 13 (2007).

[2] R. Ramesh and N. Spaldin, Nature Mater. **6**, 21 (2007).

[3] K.Wang, J.-M. Liu, and Z. Ren, Adv. Phys. **58**, 321 (2009).

[4] D. Khomskii, Physics **2**, 20 (2009).

[5] H. Katsura, N. Nagaosa, and A. V. Balatsky, Phys. Rev. Lett. **95**, 057205 (2005).

[6] C. Jia, S. Onoda, N. Nagaosa, and J. H. Han, Phys. Rev. B **74**, 224444 (2006).

[7] S. Picozzi, K. Yamauchi, B. Sanyal, I. A. Sergienko, and E. Dagotto, Phys. Rev. Lett. **99**, 227201 (2007).

[8] H. J. Xiang, E. J. Kan, Y. Zhang, M.-H. Whangbo, and X. G. Gong, Phys. Rev. Lett. **107**, 157202 (2011).

[9] I. A. Sergienko and E. Dagotto, Phys. Rev. B **73**, 094434 (2006).

[10] H. J. Xiang, S.-H. Wei, M.-H. Whangbo, and J.L.F. Da Silva, Phys. Rev. Lett. **101**, 037209 (2008).

[11] A. Malashevich and D. Vanderbilt, Phys. Rev. Lett. **101**, 037210 (2008); Phys. Rev. B **80**, 224407 (2009).

[12] X. Z. Lu, M.-H. Whangbo, S. Dong, X. G. Gong, and H. J. Xiang, Phys. Rev. Lett. **108**, 187204 (2012).

[13] J. H. Yang, Z. L. Li, X. Z. Lu, M.-H. Whangbo, S. H. Wei, X. G. Gong, and H. J. Xiang, Phys. Rev. Lett. **109**, 107203 (2012).

[14] I. A. Sergienko, C. Şen, and E. Dagotto, Phys. Rev. Lett. **97**, 227204 (2006).

[15] C. J. Wang, G.-C. Guo, and L. X. He, Phys. Rev. Lett. **99**, 177202 (2007).

[16] See supplemental material at http://link.aps.org/supplemental/.

[17] R. D. King-Smith and D. Vanderbilt, Phys. Rev. B **47**, 1651 (1993); R. Resta, Rev. Mod. Phys. **66**, 899 (1994).

[18] T. Moriya, Phys. Rev. **120**, 91 (1960).

[19] D. Dai, H. J. Xiang, and M.-H. Whangbo, J. Comput. Chem. **29**, 2187 (2008).

[20] H. J. Xiang, E. J. Kan, S.-H. Wei, M.-H. Whangbo, and X. G. Gong, Phys. Rev. B **84**, 224429 (2011).

[21] H. J. Xiang, C. Lee, H.-J. Koo, X. G. Gong, and M.-H. Whangbo, Dalton Trans. **42**, 823 (2013).

[22] X. Gonze, Phys. Rev. B **55**, 10337 (1997).

[23] S. Baroni, S. de Gironcoli, A. Dal Corso, and P. Giannozzi, Rev. Mod. Phys. **73**, 515 (2001).

[24] M. A. Lafontaine, M. Leblanc, and G. Ferey, Acta Crystallogr. Sect. C **45**, 1205 (1989).

[25] H.-J. Koo, C. Lee, M.-H. Whangbo, G. J. McIntyre and R. K. Kremer, Inorg. Chem. **50**, 3582 (2011).

[26] H. J. Xiang and M.-H. Whangbo, Phys. Rev. Lett. **99**, 257203 (2007).

[27] T. Kimura *et al.*, Nature (London) **426**, 55 (2003).

[28] J. A. Alonso, M. J. Martínez-Lope, M. T. Casais, and M. T. Fernández-Díaz, Inorg. Chem. **39**, 917 (2000).

[29] H. C. Walker, F. Fabrizi, L. Paolasini, F. de Bergevin, J. Herrero-Martin, A. T. Boothroyd, D. Prabhakaran, and D. F. McMorrow, Science **333**, 1273 (2011).

Table I. The ferroelectric polarizations (in unit of μC/m$^2$) of $LiCuVO_4$, $TbMnO_3$ and $HoMnO_3$ obtained from the DFT and the unified polarization model calculations. Here the magnetic propagation vectors used for the spin spiral are $\vec{q} = (0, 0.25, 0)$ for $LiCuVO_4$ and $\vec{q} = (0, 1/3, 0)$ for $TbMnO_3$. The ion-displacement contribution from the DFT calculations is estimated as $\vec{P}_{ion}(\text{DFT}) = \vec{P}_t(\text{DFT}) - \vec{P}_e(\text{DFT})$.

| | $LiCuVO_4$ | | $TbMnO_3$ | | $HoMnO_3$ |
|---|---|---|---|---|---|
| | *ab*-spiral | *bc*-spiral | *ab*-spiral | *bc*-spiral | E-AFM |
| $\vec{P}_e$ (model) | (-112.4, 0, 0) | (0, 0, 18.9) | (-306.7, 0, 0) | (0,0,-25.7) | $(3.47\times10^4, 0, 0)$ |
| $\vec{P}_{ion}$ (model) | (-366.9, 0, 0) | (0, 0, 187.4) | (552.2, 0, 0) | (0, 0, -477.0) | $(3.28\times10^4, 0, 0)$ |
| $\vec{P}_t$ (model) | (-479.3, 0, 0) | (0, 0, 206.3) | (245.5, 0, 0) | (0, 0, -502.7) | $(6.75\times10^4, 0, 0)$ |
| $\vec{P}_e$ (DFT) | (-103.5, 0, 0) | (0, 0, 15.7) | (-331.0, 0, 0) | (0, 0, 0.5) | $(3.19\times10^4, 0, 0)$ |
| $\vec{P}_{ion}$ (DFT) | (-492.1, 0, 0) | (0, 0, 207.3) | (462.2, 0, 0) | (0, 0, -424.5) | $(2.95\times10^4, 0, 0)$ |
| $\vec{P}_t$ (DFT) | (-595.6, 0, 0) | (0, 0, 223.0) | (131.2, 0, 0) | (0, 0, -424.0) | $(6.14\times10^4, 0, 0)$ |

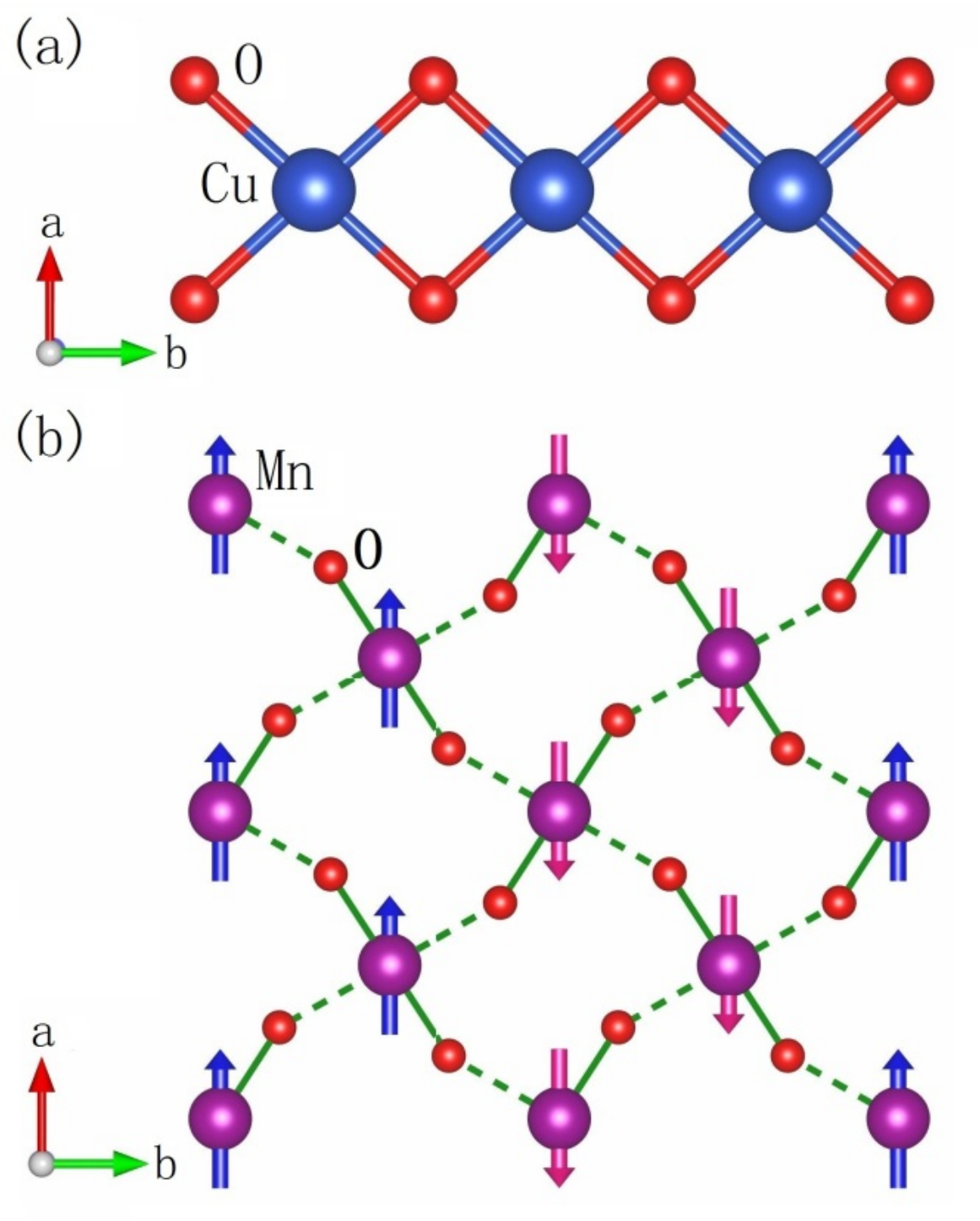

FIG. 1 (color online). (a) The one-dimensional $CuO_2$ ribbon chains made up of edge-sharing $CuO_4$ square planes. Each $Cu^{2+}$ ($d^9$, S = 1/2) ion has one singly-occupied d-level (namely, the up-spin $d_{X2-Y2}$ level, when the local X and Y axes are taken along the Cu-O bond directions). (b) An isolated $MnO_2$ plane of $RMnO_3$ (R = Tb or Ho) viewed along the *c*-axis. The solid and dashed lines denote the short and long Mn-O bonds, respectively, of an axially-elongated $MnO_6$ octahedron containing a $Mn^{3+}$ ($d^4$, S = 2) ion. With the local z axis taken along the long Mn-O bond, each $Mn^{3+}$ ion has one $e_g$ level ($d_{z2}$) occupied and one $e_g$ level ($d_{x2-yx2}$) unoccupied. The arrows indicate the directions of the Mn spins in $HoMnO_3$, which illustrate the E-AFM spin order.

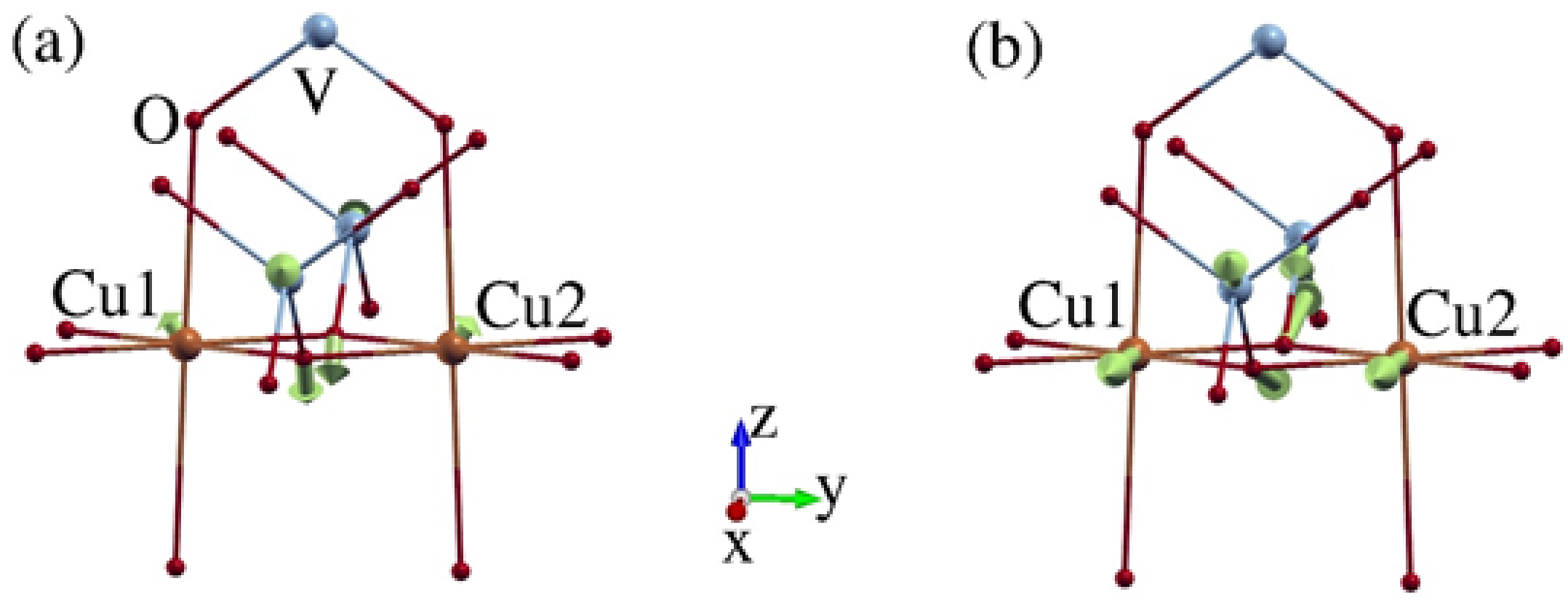

FIG. 2 (color online). The first-order derivatives associated with the DM interaction parameter between Cu1 and Cu2 in $LiCuVO_4$: (a) $\frac{\partial D_{12}^{x}}{\partial \vec{u}}$ and (b) $\frac{\partial D_{12}^{z}}{\partial \vec{u}}$. The $\frac{\partial D_{12}^{y}}{\partial \vec{u}}$ term is not shown because it is not relevant to the polarization. The coordination system is also shown. The local symmetry for the Cu dimer is $C_{2v}$.

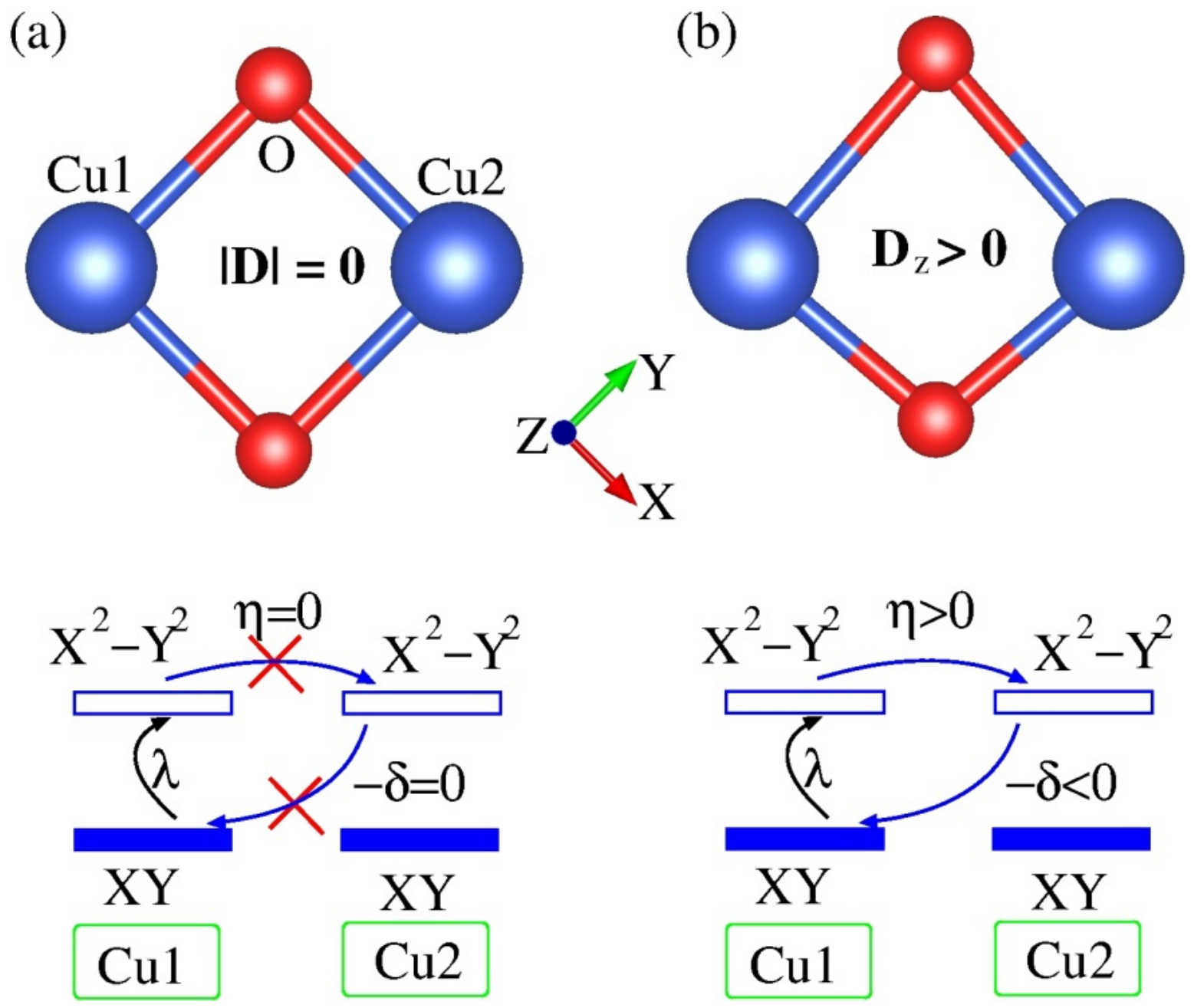

FIG. 3 (color online). DM interactions between nearest-neighbor $Cu^{2+}$ ions in $LiCuVO_4$ for the cases of (a) when the two bridging O ions are symmetric with respect to the (110) plane and (b) when the two bridging O ions move along the [-110] direction. In each case upper panel displays the structures of the $Cu_2O_2$ cluster, and the lower panel illustrates the three-step hopping process associated with the hole state $d_{X2-Y2}$ and the filled state $d_{XY}$. $\eta$ and $\delta$ represent the hopping integrals between these d states of Cu1 and Cu2, and $\lambda$ denotes the strength of the SOC at each Cu site.

Supplementary Materials for

# Unified Model of Ferroelectricity Induced by Spin Order

H. J. Xiang[1*], P. S. Wang[1], M.-H. Whangbo[2], and X. G. Gong[1]

[1] Key Laboratory of Computational Physical Sciences (Ministry of Education), State Key Laboratory of Surface Physics, and Department of Physics, Fudan University, Shanghai 200433, P. R. China

[2]Department of Chemistry, North Carolina State University, Raleigh, North Carolina 27695-8204, USA

## 1. DFT computational details

Total energy calculations are based on the DFT plus the on-site repulsion (U) method [S1] on the basis of the projector augmented wave method [S2] encoded in the Vienna ab initio simulation package [S3]. The plane-wave cutoff energy is set to 400 eV. For the calculation of electric polarization, the Berry phase method [S4] is employed. For $LiCuVO_4$, $TbMnO_3$, and $HoMnO_3$, we use the same effective U values as those adopted in previous theoretical studies [S5,S6,S7]. When calculating the electric polarization or DM interactions for $LiCuVO_4$ and $TbMnO_3$, the spin-orbit coupling effect is included.

## 2. Verification of the assumption that the electronic contribution remains almost unchanged

Our polarization model assumes that the electronic contribution ($\vec{P}_e$) in the optimized structure is the same as that in the paraelectric structure. We take $TbMnO_3$ as an example to check this assumption. Using the experimental paraelectric structure, the electronic contributions to the polarization are (-331.0, 0, 0) $\mu C/m^2$ and (0, 0, 0.5) $\mu C/m^2$ for the *ab*-spiral case and *bc*-spiral case, respectively. To calculate the electronic contribution to the total polarization in the optimized structure of a given spin-spiral order, we first relax the structure with the spin-spiral order. The total polarization of this state is defined as $\vec{P}_t$(DFT). Then the polarization [$\vec{P}_t$(DFT, FM)] of this optimized structure but with the ferromagnetic (FM) order is computed. The electronic contribution to the total polarization in the optimized structure of this spin-spiral order is then calculated as: $\vec{P}_t(\mathrm{DFT}) - \vec{P}_t(\mathrm{DFT, FM})$. Using this procedure, the electronic contributions to the polarization of the optimized structure are (-367.7, 0, 0) μC/m$^2$ and (0, 0, 26.1) μC/m$^2$ for the *ab*-spiral and *bc*-spiral cases, respectively. Therefore, for the optimized and the paraelectric structures, the electronic contributions to the polarization do not differ much (~30 μC/m$^2$).

## 3. Symmetry analysis of the general spin current model

The general spin-current term [S8] of the electronic contribution to the polarization of a spin pair ($\vec{S}_1$ and $\vec{S}_2$) can be written as:

$$\vec{\mathrm{P}} = \mathbf{M}(\vec{\mathrm{S}}_1 \times \vec{\mathrm{S}}_2), \text{ with } \mathbf{M} = \begin{bmatrix} (\mathrm{P}_{12}^{yz})_x & (\mathrm{P}_{12}^{zx})_x & (\mathrm{P}_{12}^{xy})_x \\ (\mathrm{P}_{12}^{yz})_y & (\mathrm{P}_{12}^{zx})_y & (\mathrm{P}_{12}^{xy})_y \\ (\mathrm{P}_{12}^{yz})_z & (\mathrm{P}_{12}^{zx})_z & (\mathrm{P}_{12}^{xy})_z \end{bmatrix}.$$

The local point group symmetry of the spin dimer in the crystal may place some restrictions on the form of the matrix **M**. The symmetric operations of the point group can be classified into different kinds. We have two ways to classify the spatial symmetric operations: One way is to check whether the symmetry operation **R** swaps the two magnetic ions; another way is to check whether the symmetry operation is a proper rotation [ $\det(\mathbf{R}) = 1$ ] or improper rotation [ $\det(\mathbf{R}) = -1$ ]. For a given symmetry operation **R**, we can apply the symmetry operation **R** to a spin configuration characterized by $\vec{S}_1$ and $\vec{S}_2$ to obtain a new spin configuration characterized by $\vec{S}_1'$ and $\vec{S}_2'$. The polarization for the new spin configuration can be computed in two ways: (a) Apply the rotation **R** to the polarization of the original spin configuration, namely, $\mathbf{R}\vec{\mathrm{P}} = \mathbf{RM}\ (\vec{\mathrm{S}}_1 \times \vec{\mathrm{S}}_2)$. (b) Use the general spin-current model to obtain the polarization of the new spin configuration, namely, $\mathbf{M}(\vec{S}_1' \times \vec{S}_2')$. The two approaches should lead to the same result, so that $\mathbf{RM}\ (\vec{S}_1 \times \vec{S}_2) = \mathbf{M}(\vec{S}_1' \times \vec{S}_2')$.

(i) The symmetry operation **R** does not swap the magnetic ions: If the symmetry operation **R** is a proper rotation and it does not change the positions of the magnetic ions, we have $\vec{S}_1' = \mathbf{R}\vec{S}_1$ and $\vec{S}_2' = \mathbf{R}\vec{S}_2$.

$$\mathbf{RM}\ (\vec{S}_1 \times \vec{S}_2) = \mathbf{M}(\vec{S}_1' \times \vec{S}_2') = \mathbf{M}(\mathbf{R}\vec{S}_1 \times \mathbf{R}\vec{S}_2) = \mathbf{MR}(\vec{S}_1 \times \vec{S}_2)$$
$$\therefore \mathbf{RMR}^{-1} = \mathbf{M} = \det(\mathbf{R})\mathbf{M}.$$

If the symmetry operation **R** is an improper rotation, we have $\vec{S}_1' = -\mathbf{R}\vec{S}_1$ and $\vec{S}_2' = -\mathbf{R}\vec{S}_2$ because an improper rotation **R** acts on the spin (axial vector) only through the pure rotation part ($-\mathbf{R}$).

$$\mathbf{RM}\ (\vec{S}_1 \times \vec{S}_2) = \mathbf{M}(\vec{S}_1' \times \vec{S}_2') = \mathbf{M}[(-\mathbf{R})\vec{S}_1 \times (-\mathbf{R})\vec{S}_2] = \mathbf{M}(-\mathbf{R})(\vec{S}_1 \times \vec{S}_2)$$
$$\therefore \mathbf{RMR}^{-1} = -\mathbf{M} = \det(\mathbf{R})\mathbf{M}.$$

(ii) If the symmetry operation **R** swaps the magnetic ions, we have $\vec{S}_1' = \mathbf{R}\vec{S}_2$ and $\vec{S}_2' = \mathbf{R}\vec{S}_1$ if **R** is a proper rotation:

$$\mathbf{RM}\ (\vec{S}_1 \times \vec{S}_2) = \mathbf{M}(\vec{S}_1' \times \vec{S}_2') = \mathbf{M}(\mathbf{R}\vec{S}_2 \times \mathbf{R}\vec{S}_1) = -\mathbf{MR}(\vec{S}_1 \times \vec{S}_2)$$
$$\therefore \mathbf{RMR}^{-1} = -\mathbf{M} = -\det(\mathbf{R})\mathbf{M}.$$

If the symmetry operation **R** is an improper rotation, we also have $\mathbf{RMR}^{-1} = -\det(\mathbf{R})\mathbf{M}.$

For the $Cu_2$ dimer (see Fig. 2 in the main text) in $LiCuVO_4$, it has the $C_{2v}$ point group symmetry: The symmetry elements include the two-fold rotational axis along z ($\mathbf{R_1}$), the xz mirror-plane ($\mathbf{R_2}$), and the yz mirror-plane ($\mathbf{R_3}$):

$$\mathbf{R}_1 = \begin{pmatrix} -1 & & \\ & -1 & \\ & & 1 \end{pmatrix}, \mathbf{R}_2 = \begin{pmatrix} 1 & & \\ & -1 & \\ & & 1 \end{pmatrix}, \mathbf{R}_3 = \begin{pmatrix} -1 & & \\ & 1 & \\ & & 1 \end{pmatrix}.$$

The two-fold rotational axis along z is a proper operation which swaps the two Cu ions. The xz mirror-plane is an improper operation which swaps the two Cu ions. The yz mirror-plane is an improper operation which does not swap the two Cu ions. According to the rules derived above, there are three requirements for **M** to satisfy:

1. $\mathbf{R}_1\mathbf{M}\mathbf{R}_1^{-1} = -\mathbf{M}$
2. $\mathbf{R}_2\mathbf{M}\mathbf{R}_2^{-1} = \mathbf{M}$
3. $\mathbf{R}_3\mathbf{M}\mathbf{R}_3^{-1} = -\mathbf{M}$

(a) For the first requirement $\mathbf{R}_1\mathbf{M}\mathbf{R}_1^{-1} = -\mathbf{M}$, we have:

$$\mathbf{R}_1 = \mathbf{R}_1^{-1} = \begin{pmatrix} -1 & & \\ & -1 & \\ & & 1 \end{pmatrix}, \quad \mathrm{M} = \begin{pmatrix} \mathrm{M}_{11} & \mathrm{M}_{12} & \mathrm{M}_{13} \\ \mathrm{M}_{21} & \mathrm{M}_{22} & \mathrm{M}_{23} \\ \mathrm{M}_{31} & \mathrm{M}_{32} & \mathrm{M}_{33} \end{pmatrix},$$

$$\mathbf{R}_1\mathbf{M} = \begin{pmatrix} -\mathrm{M}_{11} & -\mathrm{M}_{12} & -\mathrm{M}_{13} \\ -\mathrm{M}_{21} & -\mathrm{M}_{22} & -\mathrm{M}_{23} \\ \mathrm{M}_{31} & \mathrm{M}_{32} & \mathrm{M}_{33} \end{pmatrix}$$

$$\mathbf{R}_1\mathbf{M}\mathbf{R}_1^{-1} = \begin{pmatrix} -\mathrm{M}_{11} & -\mathrm{M}_{12} & -\mathrm{M}_{13} \\ -\mathrm{M}_{21} & -\mathrm{M}_{22} & -\mathrm{M}_{23} \\ \mathrm{M}_{31} & \mathrm{M}_{32} & \mathrm{M}_{33} \end{pmatrix} \begin{pmatrix} -1 & & \\ & -1 & \\ & & 1 \end{pmatrix} = \begin{pmatrix} \mathrm{M}_{11} & \mathrm{M}_{12} & -\mathrm{M}_{13} \\ \mathrm{M}_{21} & \mathrm{M}_{22} & -\mathrm{M}_{23} \\ -\mathrm{M}_{31} & -\mathrm{M}_{32} & \mathrm{M}_{33} \end{pmatrix}$$

$$\mathbf{R}_1\mathbf{M}\mathbf{R}_1^{-1} = -\mathbf{M}$$

$$\begin{pmatrix} \mathrm{M}_{11} & \mathrm{M}_{12} & -\mathrm{M}_{13} \\ \mathrm{M}_{21} & \mathrm{M}_{22} & -\mathrm{M}_{23} \\ -\mathrm{M}_{31} & -\mathrm{M}_{32} & \mathrm{M}_{33} \end{pmatrix} = \begin{pmatrix} -\mathrm{M}_{11} & -\mathrm{M}_{12} & -\mathrm{M}_{13} \\ -\mathrm{M}_{21} & -\mathrm{M}_{22} & -\mathrm{M}_{23} \\ -\mathrm{M}_{31} & -\mathrm{M}_{32} & -\mathrm{M}_{33} \end{pmatrix}$$

$$\therefore \mathrm{M} = \begin{pmatrix} & & \mathrm{M}_{13} \\ & & \mathrm{M}_{23} \\ \mathrm{M}_{31} & \mathrm{M}_{32} & \end{pmatrix}.$$

(b) For the second requirement $\mathbf{R}_2\mathbf{M}\mathbf{R}_2^{-1} = \mathbf{M}$, we have:

$$\mathbf{R}_2 = \mathbf{R}_2^{-1} = \begin{pmatrix} 1 & & \\ & -1 & \\ & & 1 \end{pmatrix}, \mathrm{M} = \begin{pmatrix} \mathrm{M}_{11} & \mathrm{M}_{12} & \mathrm{M}_{13} \\ \mathrm{M}_{21} & \mathrm{M}_{22} & \mathrm{M}_{23} \\ \mathrm{M}_{31} & \mathrm{M}_{32} & \mathrm{M}_{33} \end{pmatrix},$$

$$\mathbf{R}_2\mathbf{M} = \begin{pmatrix} \mathrm{M}_{11} & \mathrm{M}_{12} & \mathrm{M}_{13} \\ -\mathrm{M}_{21} & -\mathrm{M}_{22} & -\mathrm{M}_{23} \\ \mathrm{M}_{31} & \mathrm{M}_{32} & \mathrm{M}_{33} \end{pmatrix}$$

$$\mathbf{R}_2\mathbf{M}\mathbf{R}_2^{-1} = \begin{pmatrix} \mathrm{M}_{11} & \mathrm{M}_{12} & \mathrm{M}_{13} \\ -\mathrm{M}_{21} & -\mathrm{M}_{22} & -\mathrm{M}_{23} \\ \mathrm{M}_{31} & \mathrm{M}_{32} & \mathrm{M}_{33} \end{pmatrix}\begin{pmatrix} 1 & & \\ & -1 & \\ & & 1 \end{pmatrix} = \begin{pmatrix} \mathrm{M}_{11} & -\mathrm{M}_{12} & \mathrm{M}_{13} \\ -\mathrm{M}_{21} & \mathrm{M}_{22} & -\mathrm{M}_{23} \\ \mathrm{M}_{31} & -\mathrm{M}_{32} & \mathrm{M}_{33} \end{pmatrix}$$

$$\mathbf{R}_2\mathbf{M}\mathbf{R}_2^{-1} = \mathbf{M}$$

$$\begin{pmatrix} \mathrm{M}_{11} & -\mathrm{M}_{12} & \mathrm{M}_{13} \\ -\mathrm{M}_{21} & \mathrm{M}_{22} & -\mathrm{M}_{23} \\ \mathrm{M}_{31} & -\mathrm{M}_{32} & \mathrm{M}_{33} \end{pmatrix} = \begin{pmatrix} \mathrm{M}_{11} & \mathrm{M}_{12} & \mathrm{M}_{13} \\ \mathrm{M}_{21} & \mathrm{M}_{22} & \mathrm{M}_{23} \\ \mathrm{M}_{31} & \mathrm{M}_{32} & \mathrm{M}_{33} \end{pmatrix}$$

$$\therefore \mathrm{M} = \begin{pmatrix} \mathrm{M}_{11} & & \mathrm{M}_{13} \\ & \mathrm{M}_{22} & \\ \mathrm{M}_{31} & & \mathrm{M}_{33} \end{pmatrix}.$$

(c) For the first and second requirements ($\mathbf{R}_1\mathbf{M}\mathbf{R}_1^{-1} = -\mathbf{M}$ and $\mathbf{R}_2\mathbf{M}\mathbf{R}_2^{-1} = \mathbf{M}$), we have:

$$\mathrm{M} = \begin{pmatrix} & & \mathrm{M}_{13} \\ & & \\ \mathrm{M}_{31} & & \end{pmatrix}.$$

(d) The above **M** also satisfies the third requirement $\mathbf{R}_3\mathbf{M}\mathbf{R}_3^{-1} = -\mathbf{M}$.

(e) The result on $LiCuVO_4$ from the first-principles calculations is in good agreement with the symmetry analysis given above. For $TbMnO_3$, the Mn-Mn dimer has $C_1$ point group symmetry so that there is no restriction on **M**.

## 4. Estimate the DM interaction parameter for $LiCuVO_4$

The DM interaction for the Cu dimer shown in Fig. 3(b) can be evaluated using Moriya's formula [S9]:

$$
\begin{aligned}
D_{12}^{z} &= \frac{4i[b_{nn'}C_{n'n}^{z} - C_{nn'}^{z}b_{n'n}]}{U} \\
C_{n'n}^{z} &= -\frac{\lambda}{2}\left(\frac{[l_{m'n'}^{z}]^{*}}{\varepsilon_{m'}-\varepsilon_{n'}}b_{m'n} + \frac{l_{mn}^{z}}{\varepsilon_{m}-\varepsilon_{n}}b_{n'm}\right) \\
C_{nn'}^{z} &= -\frac{\lambda}{2}\left(\frac{[l_{mn}^{z}]^{*}}{\varepsilon_{m}-\varepsilon_{n}}b_{mn'} + \frac{l_{m'n'}^{z}}{\varepsilon_{m'}-\varepsilon_{n'}}b_{nm'}\right)
\end{aligned}
$$

Here $n$ and $m$ refer to the $d_{X2\text{-}Y2}$ (the hole orbital) and $d_{XY}$ orbitals of Cu1, respectively (The local coordination system XYZ is used in the TB analysis). Similarly, $n'$ and $m'$ refer to the $d_{X2\text{-}Y2}$ (the hole orbital) and $d_{XY}$ orbitals of Cu2, respectively. By symmetry, the effective hopping between the d states mediated by O 2p orbitals satisfy these relations: $b_{nn'} = b_{n'n} = \eta > 0$, $-b_{mn'} = -b_{n'm} = b_{nm'} = b_{m'n} = \delta > 0$. Since $l_{mn}^{z} = l_{m'n'}^{z} = 2i$, we have

$$
\begin{aligned}
C_{n'n}^{z} &= -\frac{\lambda}{2}\left(\frac{[l_{m'n'}^{z}]^{*}}{\varepsilon_{m'}-\varepsilon_{n'}}b_{m'n} + \frac{l_{mn}^{z}}{\varepsilon_{m}-\varepsilon_{n}}b_{n'm}\right) = \frac{2\delta\lambda i}{\Delta E} \\
C_{nn'}^{z} &= -\frac{\lambda}{2}\left(\frac{[l_{mn}^{z}]^{*}}{\varepsilon_{m}-\varepsilon_{n}}b_{mn'} + \frac{l_{m'n'}^{z}}{\varepsilon_{m'}-\varepsilon_{n'}}b_{nm'}\right) = \frac{-2\delta\lambda i}{\Delta E} \\
D_{12}^{z} &= \frac{4i[b_{nn'}C_{n'n}^{z} - C_{nn'}^{z}b_{n'n}]}{U} = \frac{4\eta i}{U}\frac{4\delta\lambda i}{\Delta E} = \frac{-16\eta\delta\lambda}{U\Delta E}
\end{aligned},
$$

where $\Delta E < 0$ is the energy difference between the $d_{X2\text{-}Y2}$ and $d_{XY}$ levels. Therefore, $D_{12}^{z} > 0$, in agreement with DFT calculations and direct TB band energy calculations.

## 5. Derivative of the symmetric exchange parameter of $HoMnO_3$

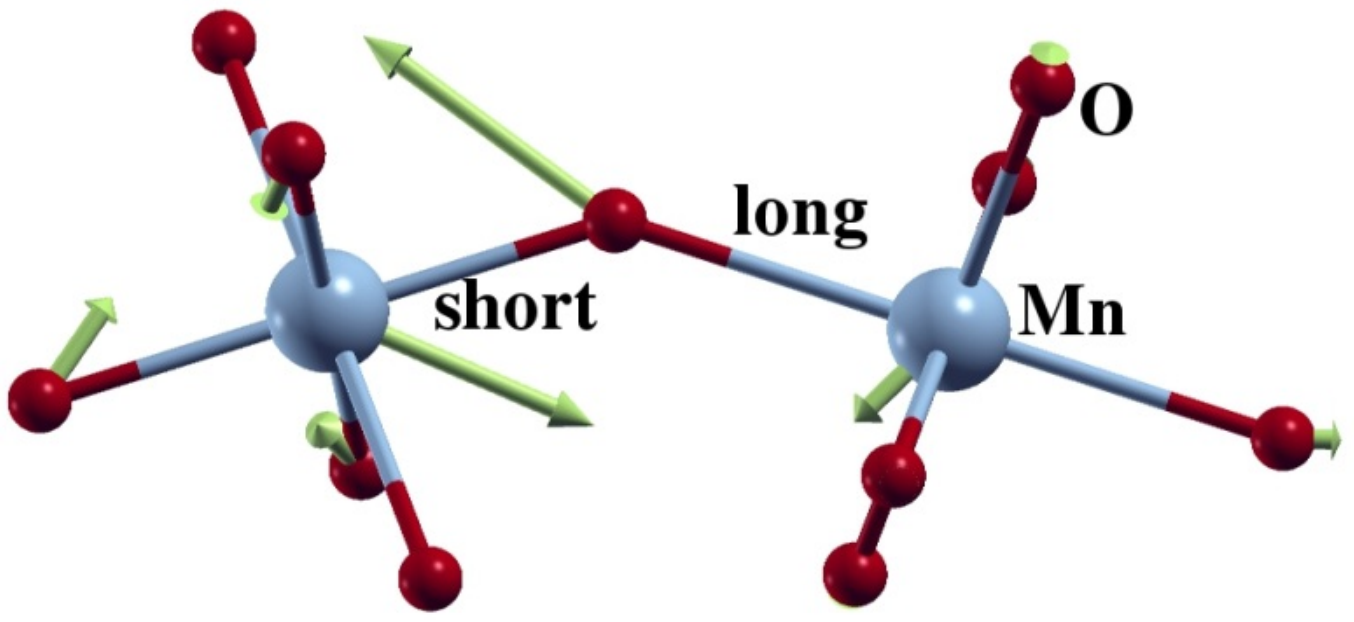

**Fig. S1.** The derivative of the in-plane nearest neighbor symmetric exchange interaction in $HoMnO_3$. The derivative vectors are almost in the *ab*-plane. These vectors show that when the Mn-O-Mn angle becomes smaller and/or the Mn-Mn distance becomes shorter, the spin exchange interaction becomes more positive (antiferromagnetic).

**References**

[S1] A. I. Liechtenstein et al., Phys. Rev. B **52**, R5467 (1995).

[S2] P. E. Blöchl, Phys. Rev. B **50**, 17953 (1994); G. Kresse and D. Joubert, Phys. Rev. B **59**, 1758 (1999).

[S3] G. Kresse and J. Furthmüller, Comput. Mater. Sci. **6**, 15 (1996); Phys. Rev. B **54**, 11169 (1996).

[S4] R. D. King-Smith and D. Vanderbilt, Phys. Rev. B **47**, 1651 (1993); R. Resta, Rev. Mod. Phys. **66**, 899 (1994).

[S5] H. J. Xiang and M. H. Whangbo, Phys. Rev. Lett. **99**, 257203 (2007).

[S6] H. J. Xiang, Su-Huai Wei, M.-H. Whangbo, and Juarez L. F. Da Silva, Phys. Rev. Lett. **101**, 037209 (2008).

[S7] S. Picozzi, K. Yamauchi, B. Sanyal, I. A. Sergienko, and E. Dagotto, Phys. Rev.

Lett. **99**, 227201 (2007).

[S8] H. J. Xiang, E. J. Kan, Y. Zhang, M.-H. Whangbo, and X. G. Gong, Phys. Rev. Lett. **107**, 157202 (2011)

[S9] T. Moriya, Phys. Rev. **120**, 91 (1960).